\documentclass[a4paper]{jpconf}
\usepackage{multicol}
\usepackage{graphicx}
\usepackage{amsmath,amssymb}
\usepackage{color}

\definecolor{red}{rgb}{1,0,0}
\definecolor{green}{rgb}{0,1,0}
\definecolor{blue}{rgb}{0,0,1}

\begin{document}
\title{Optimal models of extreme volume-prices are time-dependent}

\author{Paulo Rocha$^1$, Frank Raischel$^2$, Jo\~ao Pedro Boto$^1$ and 
        Pedro G.~Lind$^3$} 
  \address{%
  $^{1}$ Centro de Matem\'atica e Aplica\c{c}\~oes Fundamentais
  Avenida Professor Gama Pinto, 2
  1649-003 Lisboa, Portugal\\}  
  \address{%
  $^2$Instituto Dom Luiz, CGUL,
  University of Lisbon,
  1749-016 Lisbon, Portugal\\}
  \address{%
  $^3$ForWind and Institute of Physics,
  University of Oldenburg,
  DE-26111 Oldenburg, Germany\\
  }  
  \ead{paulorocha99@hotmail.com, raischel@cii.fc.ul.pt, boto@ptmat.fc.ul.pt, pedro.g.lind@forwind.de}
  
\begin{abstract}
We present evidence that the best model for empirical
volume-price distributions is not always the same and
it strongly depends in (i) the region of the volume-price
spectrum that one wants to model and (ii) the period in time
that is being modelled.
To show these two features we analyze stocks of the New York
stock market with four different models: $\Gamma$, $\Gamma$-inverse, 
log-normal, and Weibull distributions.
To evaluate the accuracy of each model we use standard 
relative deviations as well as the Kullback-Leibler distance
and introduce an additional distance particularly suited
to evaluate how accurate are the models for the distribution
tails (large volume-price).
Finally we put our findings in perspective and discuss how
they can be extended to other situations in finance engineering.
\end{abstract}

\section{Introduction}

It has been acknowledge in financial distributions that the statistics 
of extreme events, leading to heavy tails, as well has correlations 
between noise sources and other components need to be taken into account 
if one pretends to establish a model that describe accurately this 
dynamical system. In this paper we are interested in the study of extreme 
events present in financial markets distributions. As study subject we 
shall consider the New York stock market (NYSM). We focus here on the 
evolution of volume-price, i.e.~on changes in capitalization, which 
should have more the character of a conserved quantity than the price 
per se. 
Having access to the overall distribution of volume-price 
provides information about the entire capital traded in the market.
However, extreme events focus in what happens on the right tails of
such distributions, the region comprising the highest volume-prices
occurring in a given market.

In this paper we will address the problem of modelling such tails
at high volume-prices in comparison with models for the full 
volume-price distribution.

Our data base is composed by $\sim 2000$ listed shares 
extracted from the NYSM and freely available at 
{\tt http://finance.yahoo.com/}.
The data has a sample frequency of $0.1$ min$^{-1}$, starting in 
January 27th 2011 and ending in April 6th 2014, with a total of
976 days ($\sim10^5$ data points). 

Using this data we build the volume-price distribution of the 
$\sim 2000$ listed companies every $10$ minutes.
We define the volume-price $s=pV$ as the product of the trading 
volume $V$ and the last trade price $p$. More details, concerning
the data mining and collection can be found in Ref.~\cite{SEoSM}.

We start in Sec.~\ref{sec:models} by studying the behaviour of 
volume-price distribution and introducing four different typical 
models in finance to fit the empirical data\cite{silvio13,SEoSM}. 
In Sec.~\ref{sec:errors}, we present the error analysis of each 
models, using standard relative deviations and more sophisticated
measures, namely the Kullback-Leibler distance. This measure
is particularly suited to evaluate how close the (non-linear) 
models fit is from the empirical distributions.
Section \ref{sec:conclusions} concludes the paper.

\section{Four models for volume-price distributions}
\label{sec:models}

\begin{figure}[t]
  \centering
  \includegraphics[width=0.7 \linewidth]{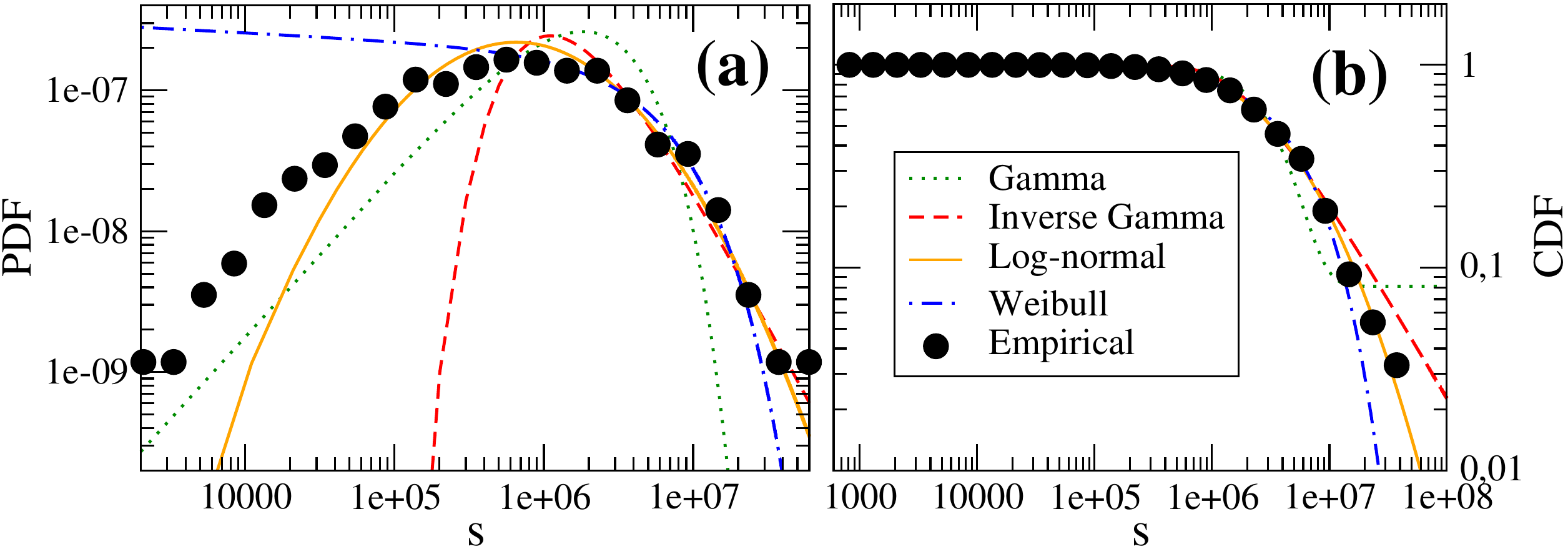}
  \caption{\protect{
         \textbf{(a)} Probability density function and 
         \textbf{(b)} cumulative density function given 
         by the  four different distributions: 
         $\Gamma$ (Eq.~\ref{gamma}), 
         inverse $\Gamma$ (Eq.~\ref{invgamma}), 
         log-normal (Eq.~\ref{lognorm}), 
         and Weibull (Eq.~\ref{weibull}).}
  }
  \label{fig01}
\end{figure}

We will consider four well-known\cite{silvio13} bi-parametric 
distributions to fit the empirical volume-price distribution, namely 
the $\Gamma-$distribution:
  \begin{equation}
  p_\Gamma(s)= \frac{s^{\phi-1}}{\theta^\phi\Gamma[\phi]}e^{-\frac{s}{\theta}}
  \label{gamma}
  \end{equation}
the inverse $\Gamma-$distribution:
  \begin{equation}
  p_{1/\Gamma}(s)= \frac{\theta^\phi}{\Gamma[\phi]}s^{-\phi-1}e^{-\frac{\theta}{s}}\
  \label{invgamma}
  \end{equation}
the Log-normal distribution:
   \begin{equation}
     p_{ln}(s)= \frac{1}{\sqrt{2\pi}\theta s}e^{-\frac{(\log s-\phi)^2}{2\theta^2}}
   \end{equation}
  \label{lognorm}
the Weibull distribution
   \begin{equation}
     p_{w}(s)= \frac{\phi}{\theta^\phi}s^{\phi-1}e^{-\left(\frac{s}{\theta}\right)^\phi\vspace{0.5cm}}
   \label{weibull}
   \end{equation}

Using the least square scheme, we fit the empirical cumulative density 
function (CDF) with each one of the four models above.
Figures \ref{fig01}a and \ref{fig01}b show respectively the 
probability and corresponding cumulative density function of each
model (lines) that fit the empirical distribution (bullets) at one
particular $10$-minute time-step.

While the log-normal and the $\Gamma$-distributions are the best ones for
the low range of volume-price values, all four models except $\Gamma$ are
closer to the right tail comprising the largest volume-prices observed.
Therefore, one should not expect to obtain always the same optimal model,
i.e.~the smallest deviations or error measures are not observed always for 
the same model.

\begin{figure}[t]
  \centering
  \includegraphics[width=0.7 \linewidth]{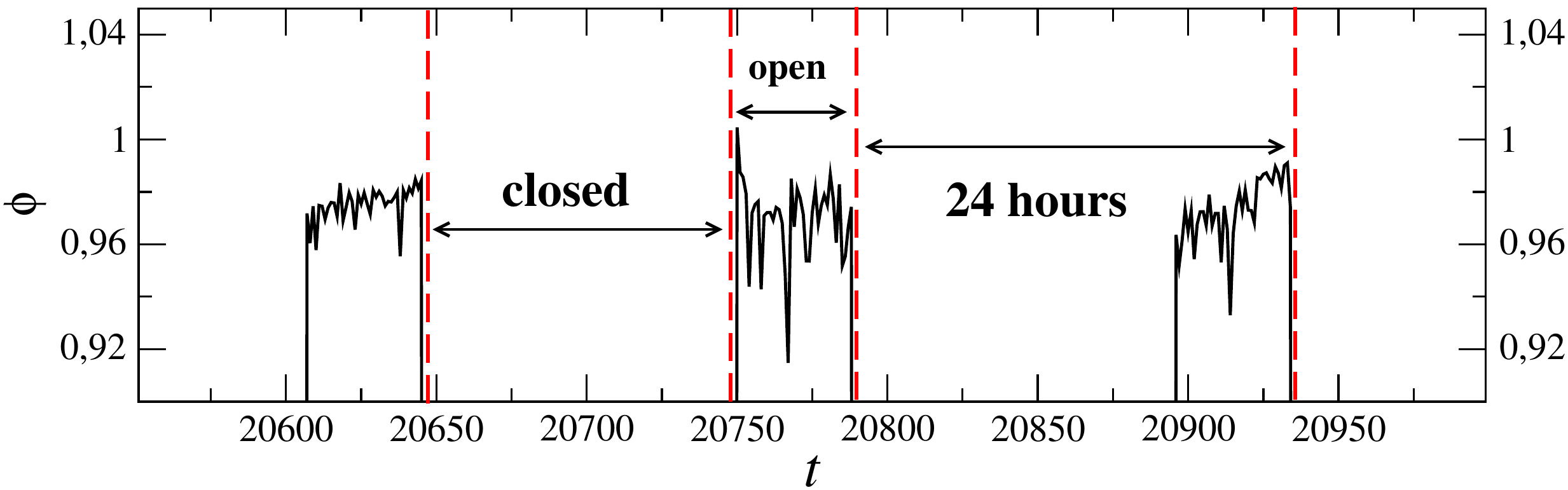}
  \caption{\protect
         Illustrative time series of the parameter $\phi$ for inverse $\Gamma-$distribution 
         (Eq.~\ref{invgamma}) during approximately three days.
         Similar plots are obtain for all parameters in the
         four models here considered.}
  \label{fig02}
\end{figure}

In Fig.~\ref{fig02} we show the typical evolution of one parameter during
a short time-interval ($\sim 3$ days). Here one plots the parameter $\phi$ 
of the inverse $\Gamma$-distribution. Similar (qualitative) evolutions are observed with other 
parameters above.
\begin{figure}[t]
  \centering
  \includegraphics[width=0.7 \linewidth]{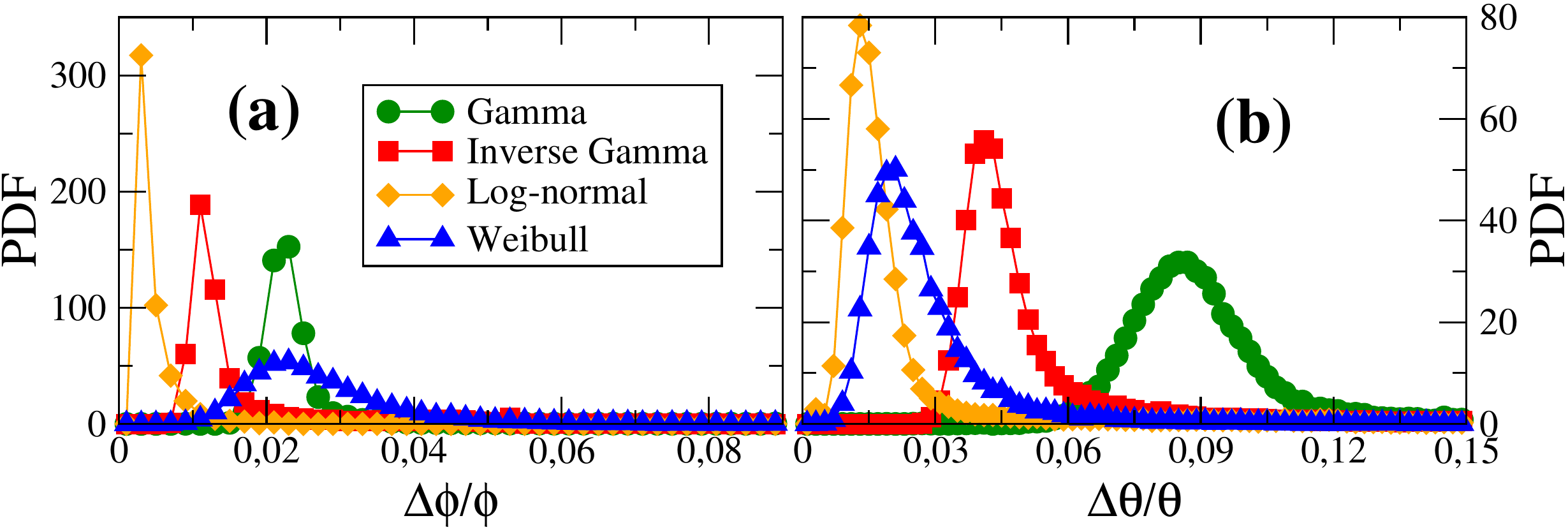}
  \caption{\protect
         Probability density function of the resulting relative 
         errors
         {\bf (a)} $\Delta \phi$ and {\bf (b)} and $\Delta \theta$,
         corresponding to the fitting parameters $\phi$ and $\theta$
         respectively.}
  \label{fig03}
\end{figure}

\section{What is the best model for our data?}
\label{sec:errors}

To evaluate how accurate a model is, we first consider the relative error,
$\Delta \phi$ and $\Delta \theta$, of each parameter value, $\phi$ and
$\theta$ respectively.
In Fig.~\ref{fig03}a and Fig.~\ref{fig03}b one plots the distributions of 
the observed relative errors.

For both parameters the smallest relative deviation is observed for the
log-normal distribution. The inverse $\Gamma$-distribution shows also
acceptable deviations. The other two models are not as good as the log-normal
and $\Gamma$-distributions. In a previous work, where the evolution of
the mean volume-price $\langle s \rangle$ was considered separately and 
the models were used to fit the distribution of the normalized 
volume-price, $s/\langle s \rangle$, the optimal model according to
relative deviations was only the inverse $\Gamma$-distribution.

\begin{figure}[htb]
  \centering
  \includegraphics[width=0.7 \linewidth]{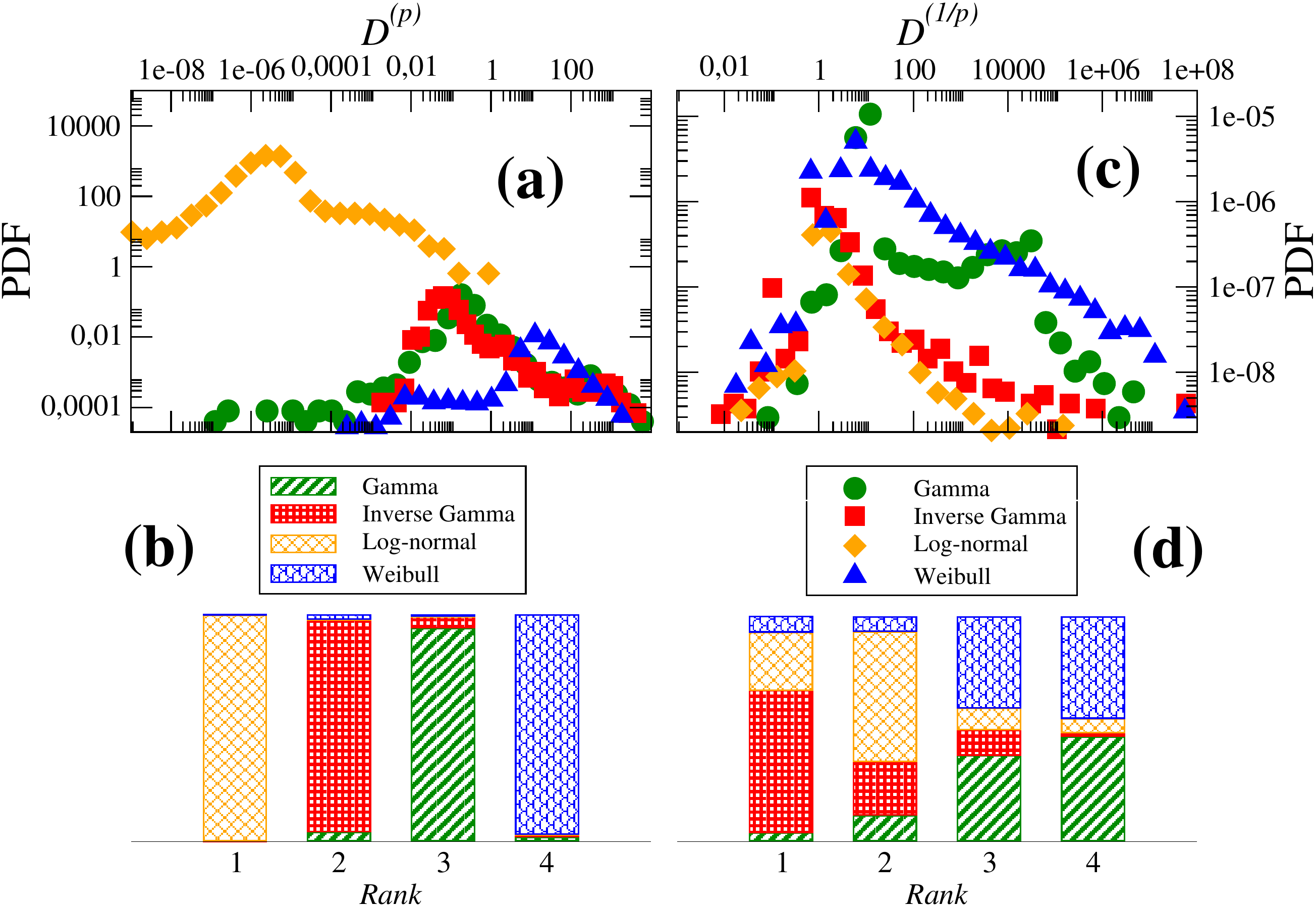}
  \caption{\protect
   {\bf (a)} PDF of the Kullback-Leibler distance $D^{(P)}$ test 
             for the full spectrum of the volume-price.   
   {\bf (b)} Percentage of accuracy rankings for each model, using the 
             Kullback-Leibler distance $D^{(P)}$. 
   A model with rank $1$ is more accurate then a model with rank $2$.
   {\bf (c)} PDF of the tail Kullback-Leibler distance $D^{(1/P)}$ 
             (see text) using only the values of $s$ larger the median 
             of the distribution.
   {\bf (d)} Percentage of accuracy rankings for each model, using the 
             tail Kullback-Leibler distance $D^{(1/P)}$.}
  \label{fig04}
\end{figure}

The relative deviations do no take into account the importance of the
volume-price spectrum in the distribution, weighted by the probability
density function or another density function.
To weight each value in the volume-price spectrum according to some
density function we introduce here the {\it generalized} Kullback-Leibler
distance:
\begin{equation}
D^{(F)}(P||Q)=\sum_i \ln\left(\left\lvert \frac{P(i)}{Q(i)} \right\rvert \right)
F(i)\Delta x ,
\label{kbdist}
\end{equation}
where $Q(i)$ is the empirical distribution, $P(i)$ is the modelled
PDF and $F(i)$ is a weighting function.
For $F(i)=P(i)$ one obtains the standard Kullback-Leibler distance\cite{kullback},
where the logarithmic deviations are heavier weighted in the central
region of the distribution. 
Figure \ref{fig04}a shows the distribution of the $D^{(P)}$ values obtained
when considering each one of the four models. Once again one observes that
the log-normal distribution is the one yielding smaller deviations.

Since the $D^{(P)}$ distributions overlap, one may argue that the log-normal
distribution might not be {\it always} the best model. To address this question
we plot in \ref{fig04}b 
the ranking ordering all four models in their accuracy for each time step. 
Indeed, almost always the log-normal is the best model, followed by the
inverse $\Gamma$-distribution.

For financial purposes, volume-price values are not equally important.
For instance, a deviation from the observed distribution in the
region of small volume-prices result in a smaller fluctuation of the
amount of transactions than in the region of largest values, where the
risk is the highest and therefore should be more accurately fitted.

To weight the largest volume-prices we first consider only the region
of the distribution for $s$ larger than the median and then consider in
Eq.~(\ref{kbdist}) $F(i)=1/P(i)$. In this way, the largest values of the
volume-price, i.e.~those for which $P(i)$ is the smallest will be 
weighted heavier than the others. Figures \ref{fig04}c and \ref{fig04}d
show the distance distributions for this case and the corresponding
rankings respectively. 

Interestingly, not only the best model is the inverse 
$\Gamma$-distribution but the dominance of one single model in each 
rank is not strong as when considering the full distributions.
This indicates that in NYSM the best model of the volume-price tail
distribution is most probably the inverse $\Gamma$-distribution but 
the probability that another model, most probably the log-normal, 
is observed as the best one is significative.

\section{Conclusions}
\label{sec:conclusions}

In this paper we analyzed New York stock market volume-price distributions. 
By testing four different  models, commonly applied in finance, to the 
empirical volume-price distributions, we found evidence that the best model 
is not always the same. 

In particular, we show that it strongly depends on the region of the spectrum 
that one wants to model, being the log-normal the best model for low values 
of $s$ and the inverse $\Gamma$ for the tail distributions.
Moreover, the best model seems to change depending on the period of time 
that the distribution is being modelled. 

All in all, our findings put in question common assumptions used when 
analyzing finance data. Namely,
to assume that volume-price distributions are log-normal\cite{Osborne} seems
to be valid
only under particular restrictions and it seems not to be
the optimal model when addressing extreme events. Furthermore,
we provided evidence to claim that the non-stationary character of 
finance data is reflected not only in the fluctuations of the
statistical moments defining the empirical distributions of 
observed values, but also in the functional shape of those 
distributions.  

For further investigation, one can 
ask what is the reason behind this time dependency. 
A good study of this strange behaviour can eventually enable one 
to improve measures of risk and to provide additional insight in risk 
management.

\section*{Acknowledgments}

The authors thank Jo\~ao P.~da Cruz for useful discussions and
{\it Funda\c{c}\~ao para a Ci\^encia e a Tecnologia}, 
{\it Deutscher Akademischer Auslandsdienst},
{\it Centro de Matem\'atica e Aplica\c{c}\~oes Fundamentais} and
German Environment Ministry for
financial support.


\end{document}